\documentclass[12pt]{article}
\usepackage{graphicx}

\newcommand{\text}{\rm}

\begin{document}

\title{\textbf{A study of the Gribov copies in linear covariant gauges in Euclidean
Yang-Mills theories }}
\author{R.F. Sobreiro\thanks{%
sobreiro@dft.if.uerj.br}, S.P. Sorella\thanks{%
sorella@uerj.br}{\ }\footnote{Work supported by FAPERJ, Funda{\c
c}{\~a}o de Amparo {\`a} Pesquisa do Estado do Rio de Janeiro,
under the program {\it Cientista do Nosso Estado}, E-26/151.947/2004.} \\
{\small {\textit{Departamento de F\'{\i}sica Te\'orica,}}}\\
{\small {\textit{Instituto de F\'{\i}sica, Universidade do Estado do Rio de
Janeiro,}}}\\
{\small {\textit{\ Rua S\~{a}o Francisco Xavier 524, 20550-013 Maracan\~{a},
}}}{\small {\textit{Rio de Janeiro, Brazil.}}}}
\date{}
\maketitle

\begin{abstract}
The Gribov copies and their consequences on the infrared behavior
of the gluon propagator are investigated in Euclidean Yang-Mills
theories quantized in linear covariant gauges. Considering small
values of the gauge parameter, it turns out that the transverse
component of the gluon propagator is suppressed, while its
longitudinal part is left unchanged. A Green function,
$\mathcal{G}_{tr}(k)$, which displays infrared enhancement and
which reduces to the ghost propagator in the Landau gauge is
identified. The inclusion of the dimension two gluon condensate
$\langle A_{\mu}^2 \rangle$ is also considered. In this case, the
transverse component of the gluon propagator and the Green
function $\mathcal{G}_{tr}(k)$ remain suppressed and enhanced,
respectively. Moreover, the longitudinal part of the gluon
propagator becomes suppressed. A comparison with the results
obtained from the studies of the Schwinger-Dyson equations and
from lattice simulations is provided.
\end{abstract}

\newpage\ \makeatother

\renewcommand{\theequation}{\thesection.\arabic{equation}}

\section{Introduction}

Gribov ambiguities \cite{Gribov:1977wm} and their relevance for
the nonperturbative aspects of Euclidean Yang-Mills theories have
witnessed growing interest in recent years. These ambiguities,
affecting the Faddeev-Popov quantization formula, deeply modify
the infrared behavior of Yang-Mills theories, as one learns from
the large amount of results obtained in the Landau gauge
\cite{Zwanziger:1982na,Zwanziger:1988jt,Zwanziger:1989mf,Dell'Antonio:1989jn,Dell'Antonio:1991xt,Zwanziger:1992qr}
as well as in the Coulomb gauge
\cite{Cucchieri:1996ja,Zwanziger:2002sh,Greensite:2004ke,Feuchter:2004mk,Reinhardt:2004mm}.
\newline
\newline
As pointed out in \cite{Singer:1978dk}, the existence of these
ambiguities is due to the lack of a globally well defined
gauge-fixing procedure. Among the class of covariant gauges, the
Gribov ambiguities have been much
investigated in the Landau gauge, where the gauge field is transverse, $%
\partial _{\mu }A_{\mu }^{a}=0$. This property plays an
important role here. It ensures that the Faddeev-Popov operator, $\mathcal{M}%
^{ab}(A)=-\partial _{\mu }\left( \partial _{\mu }\delta
^{ab}-gf^{abc}A_{\mu }^{c}\right) $, is hermitian,
$\mathcal{M=M}^{\dagger }$. Its eigenvalues are thus real.
Concerning now the quantization of Yang-Mills theories in the
Landau gauge, it turns out that, as a consequence of the existence
of Gribov copies, the domain of integration in the Feynman path
integral has to be restricted to the so called Gribov region
$\Omega $
\cite{Gribov:1977wm,Zwanziger:1982na,Zwanziger:1988jt,Zwanziger:1989mf,Dell'Antonio:1989jn,
Dell'Antonio:1991xt,Zwanziger:1992qr}, which is the set of all
transverse fields for which the Faddeev-Popov operator is positive
definite, \textit{i.e. }$\Omega =\left\{ A_{\mu }^{a},\;\partial
_{\mu }A_{\mu }^{a}=0,\mathcal{M}^{ab}(A)=-\partial _{\mu }\left(
\partial _{\mu
}\delta ^{ab}-gf^{abc}A_{\mu }^{c}\right) >0\right\} $. The boundary $%
\partial \Omega ,\,$where the first vanishing eigenvalue of the
Faddeev-Popov operator appears, is called the first Gribov horizon. For the
partition function of Yang-Mills theories in the Landau gauge, one has
\begin{eqnarray}
\mathcal{Z} &=&\int_{\Omega }DA\delta (\partial A)\left( \det \left(
-\partial ^{2}\delta ^{ab}+gf^{abc}A_{\mu }^{c}\partial _{\mu }\right)
\right) e^{-\frac{1}{4}\int d^{4}xF_{\mu \nu }^{a}F_{\mu \nu }^{a}}\;
\nonumber \\
&=&\int_{\Omega }DAD\overline{c}Dc\delta (\partial
A)e^{-\frac{1}{4}\int d^{4}xF_{\mu \nu }^{a}F_{\mu \nu }^{a}+\int
d^{4}x\;\overline{c}^{a}\partial_\mu\left(
\partial_\mu \delta ^{ab}-gf^{abc}A_{\mu }^{c}\right)
c^{b}}\;.\;  \label{zl}
\end{eqnarray}
The restriction of the domain of integration to the region $\Omega
$ has important consequences on the infrared behavior of the gluon
and ghost propagators
\cite{Gribov:1977wm,Zwanziger:1982na,Zwanziger:1988jt,Zwanziger:1989mf,Dell'Antonio:1989jn,Zwanziger:1992qr}.
More precisely, in the Landau gauge, the gluon propagator $%
\left\langle A_{\mu }^{a}(k)A_{\nu }^{b}(-k)\right\rangle $ is
found to be suppressed in the infrared, while the ghost propagator
$\left\langle \overline{c}^{a}(k)c^{b}(-k)\right\rangle $ is
enhanced, according to
\begin{equation}
\left\langle A_{\mu }^{a}(k)A_{\nu }^{b}(-k)\right\rangle =\delta
^{ab}\left( \delta _{\mu \nu }-\frac{k_{\mu }k_{\nu }}{k^{2}}\right) \frac{%
k^{2}}{k^{4}+\gamma ^{4}}\;,  \label{la}
\end{equation}
and
\begin{eqnarray}
\mathcal{G}_{gh}(k) &=&\frac{1}{N^{2}-1}\sum_{ab} \delta ^{ab}\left\langle \overline{c}%
^{a}(k)c^{b}(-k)\right\rangle \;,  \nonumber \\
\mathcal{G}_{gh}(k)_{k\rightarrow 0} &\sim &\frac{1}{k^{4}}\;.
\label{lg}
\end{eqnarray}
The Gribov parameter $\gamma $ in eq.$\left( \ref{la}\right) $ has the
dimension of a mass and is determined by a gap equation which, to the first
order approximation, reads
\begin{equation}
\frac{3Ng^{2}}{4}\int \frac{d^{4}k}{\left( 2\pi \right) ^{4}}\frac{1}{%
k^{4}+\gamma ^{4}}=1\;.  \label{gapl}
\end{equation}
From equation $\left( \ref{lg}\right) $, one sees that the
infrared behavior
of the ghost propagator is more singular than the perturbative behavior $%
1/k^{2}$. This infrared enhancement is known as the
Gribov-Zwanziger horizon condition
\cite{Gribov:1977wm,Zwanziger:1989mf,Zwanziger:1992qr}, generally stated as
$\lim_{k\rightarrow 0}\left( k^{2}\mathcal{G}_{gh}%
(k)\right) ^{-1}=0$. Remarkably, this behavior of the gluon and
ghost propagators in the Landau gauge has received many
confirmations from lattice simulations
\cite{Marenzoni:1994ap,Suman:1995zg,Cucchieri:1999sz,Bonnet:2001uh,Langfeld:2001cz,
Cucchieri:2003di,Bloch:2003sk,Furui:2003jr,Silva:2004bv} as well
as from recent studies of the Schwinger-Dyson equations
\cite{vonSmekal:1997is,Atkinson:1997tu,Atkinson:1998zc,Watson:2001yv,Zwanziger:2001kw,Lerche:2002ep,Aguilar:2004sw}.
\newline
\newline
Several results have been established about the Gribov region $%
\Omega $. It has been proven that $\Omega $ is convex and bounded
in every direction \cite{Zwanziger:1982na,Dell'Antonio:1989jn}.
Moreover, every gauge orbit passes inside $\Omega $
\cite{Dell'Antonio:1991xt}. The latter result is deeply related to
the possibility of introducing an auxiliary functional,
$\mathcal{F}(A)=\int d^{4}xA_{\mu }^{a}A_{\mu }^{a}$, whose
minimization with respect to the gauge transformations provides a
characterization of the region $\Omega $
\cite{Dell'Antonio:1991xt,Semenov}. It can be checked in fact that
the Gribov region $\Omega $ can be identified as the set of all
relative minima of the functional $\mathcal{F}(A)$. It should also
be mentioned here that the region $\Omega $ is not free from
Gribov copies, \textit{i.e. }Gribov copies still exist inside
$\Omega $ \cite{Dell'Antonio:1991xt,Semenov,vanBaal:1991zw}. To
avoid the presence of these additional copies, a further
restriction to a smaller region $\Lambda $, known as the
fundamental modular region, should be implemented
\cite{Dell'Antonio:1991xt,Semenov,vanBaal:1991zw}. The region
$\Lambda $ is contained in the Gribov region $\Omega $, being
defined as the set of all absolute minima of the auxiliary functional $%
\mathcal{F}(A)$. However, it is difficult to give an explicit
description of $\Lambda $. Recently, it has been argued in
\cite{Zwanziger:2003cf} that the additional copies existing inside
$\Omega $ have no influence on the expectation values, so that
averages calculated over $\Lambda $ or $\Omega $ are expected to
give the same result.
\newline
\newline
Besides the Landau gauge, the effects of the Gribov copies on the
infrared behavior of Yang-Mills theories have been studied to a
great extent in the noncovariant Coulomb gauge
\cite{Cucchieri:1996ja,Zwanziger:2002sh,Greensite:2004ke,Feuchter:2004mk,Reinhardt:2004mm},
$\partial _{i}A_{i}^{a}=0$, $i=1,2,3$. In particular, the Coulomb
gauge allows for a direct study of the confining properties of the
static potential $V(r)$ between an external quark pair at spatial
separation $r$. It turns out that $V(r)$ can be accessed by means
of
the $44$-component, $\left\langle A_{4}^{a}(\overrightarrow{x}%
,t)A_{4}^{b}(0)\right\rangle $, of the gluon propagator. Moreover,
in analogy with the Landau gauge, the spatial components of the
gluon propagator, $\left\langle
A_{i}^{a}(k)A_{j}^{b}(-k)\right\rangle $, are found to be
suppressed in the infrared, a feature confirmed by lattice
simulations \cite{Cucchieri:2000gu,Cucchieri:2000kw}. \newline
\newline Concerning now other covariant gauges, although the
presence of the Gribov copies is certainly to be expected
\cite{Singer:1978dk}, as explicitly confirmed by a recent study of
the zero modes of the Faddeev-Popov operator in the maximal
Abelian gauge \cite{Bruckmann:2000xd}, very little is known about
their consequences on the infrared behavior of the gluon and ghost
propagators and, more generally, on the whole Yang-Mills theory.
The need for such an investigation is motivated by the increasing
belief that the Gribov copies might have a crucial role for the
infrared region of Yang-Mills theories as well as for color
confinement. It would be desirable thus to improve our
understanding on how the Gribov copies manifest themselves in
different gauges and how they modify the infrared behavior of the
theory. \newline
\newline
This work aims at studying this issue in the class of covariant
linear gauges, $\partial _{\mu }A_{\mu }^{a}=-\alpha b^{a}$, where
$\alpha $ is the gauge parameter and $b^{a}$ is the Lagrange
multiplier. The task is far from being trivial. Many features of
the Landau gauge are lost for a
generic nonvanishing $\alpha $. The Faddeev-Popov operator, $\mathcal{M}%
^{ab}(A)=-\partial _{\mu }\left( \partial _{\mu }\delta
^{ab}-gf^{abc}A_{\mu }^{c}\right) $, is now not hermitian.
Moreover, a suitable minimizing functional in these gauges is not
yet at our disposal. As a consequence, the identification of the
region to which the domain of integration in the path-integral
should be restricted becomes difficult to be handled. All this
makes a complete treatment of the Gribov copies in linear
covariant gauges far beyond our present capabilities.
Nevertheless, we shall be able to establish some preliminary
results which enable us to obtain a characterization of the
infrared behavior of the gluon propagator, at least for small
values of the gauge parameter $\alpha $. Considering in fact small
values of the gauge parameter $\alpha $, will allow us to stay
close to the Landau gauge fixing. The present covariant gauge can
be considered thus as a kind of deformation of the Landau gauge,
whose results have to be recovered in the limit $\alpha
\rightarrow 0$, thereby providing a useful consistency check.
\newline
\newline
The output of our findings can be summarized as follows. As in the case of
the Landau gauge, the transverse component of the gluon propagator turns out
to be suppressed in the infrared. Moreover, its longitudinal part remains
unchanged. \newline
\newline
Concerning now the behavior of the ghost fields in linear covariant gauges,
it turns out that, instead of the ghost propagator, the Green function which
is enhanced in the infrared is given by the quantity $\mathcal{G}_{tr}(k)$,
defined as
\begin{eqnarray}
\mathcal{G}_{tr}(k) &=&\frac{1}{N^{2}-1}\sum_{ab} \delta ^{ab}\mathcal{G}%
_{tr}^{ab}(k)\;,  \nonumber \\
\mathcal{G}_{tr}^{ab}(k) &=&\left\langle k\left| \left( \mathcal{M}%
^{-1}(A^{T})\right) ^{ab}\right| k\right\rangle \;,  \label{gtr}
\end{eqnarray}
where
\begin{equation}
A_{\mu}^{aT}=\left( \delta _{\mu \nu }-\frac{\partial _{\mu }\partial _{\nu }}{%
\partial ^{2}}\right) A_{\nu }^{a} \;, \label{m2}
\end{equation}
is the transverse component of the gauge field and $\left(
\mathcal{M}^{-1}(A^{T})\right) ^{ab}$ stands for the inverse of
the Faddeev-Popov operator $\mathcal{M}^{ab}(A^T)$,
\begin{equation}
\mathcal{M}^{ab}(A^T)=-\partial _{\mu }\left(
\partial _{\mu }\delta ^{ab}-gf^{abc}A_{\mu }^{Tc}\right) \;.
\label{m1}
\end{equation}
The Green function $\mathcal{G}_{tr}(k)$ is found to obey the
Gribov-Zwanziger horizon condition, \textit{i.e.}
$\lim_{k\rightarrow 0}\left( k^{2}\mathcal{G}_{tr}(k)\right)
^{-1}=0$. It should be remarked that $\mathcal{G}_{tr}(k)$ does
not coincide with the ghost propagator for a generic value of the
gauge parameter $\alpha $. However, $\mathcal{G}_{tr}(k) $ reduces
to the ghost two-point function for vanishing $\alpha $, so that
our results turn out to coincide with those of the Landau gauge in
the limit $\alpha \rightarrow 0$. \newline
\newline
The infrared behavior of the gluon propagator and of
$\mathcal{G}_{tr}(k)$ will be investigated also in the presence of
the dimension two gluon condensate $\left\langle A_{\mu
}^{a}A_{\mu }^{a}\right\rangle $. A detailed study of this
condensate in linear covariant gauges can be found in
\cite{Dudal:2003by}. In the presence of $\left\langle A_{\mu
}^{a}A_{\mu }^{a}\right\rangle $, the infrared suppression of the
transverse component of the gluon propagator is enforced.
Furthermore, its longitudinal component turns out to be suppressed
as well. Concerning the Green function $\mathcal{G}_{tr}(k)$, its
infrared enhancement is not modified by the condensate
$\left\langle A_{\mu }^{a}A_{\mu }^{a}\right\rangle $.  \newline
\newline
The work is organized as follows. In Sect.2 a few properties of
the Gribov copies in the linear covariant gauges, and for small
values of the gauge parameter $\alpha $, are presented. In Sect.3
the infrared behavior of the gluon propagator, of the Green
function $\mathcal{G}_{tr}(k)$ and of the ghost propagator is
worked out. Sect.4 accounts for the inclusion of the gluon
condensate $\left\langle A_{\mu }^{a}A_{\mu }^{a}\right\rangle $.
Sect.5 is devoted to a comparison of our results with those
obtained from the analysis of the Schwinger-Dyson equations and
from lattice simulations. In Sect.6 we present our conclusion.

\section{A few properties of the Gribov copies in the linear covariant gauges
}

In this section we shall discuss a few properties of the Gribov
copies in the linear covariant gauges. Let us begin by considering
the expression of the Euclidean Yang-Mills action quantized in the
linear covariant $\alpha -$gauges, namely
\begin{equation}
S_{YM}+S_{gf}=\frac{1}{4}\int d^{4}xF_{\mu \nu }^{a}F_{\mu \nu }^{a}-\int
d^{4}x\left( b^{a}\partial A^{a}+\frac{\alpha }{2}b^{a}b^{a}+\overline{c}%
^{a}\partial _{\mu }\left( \partial _{\mu }\delta ^{ab}-gf^{abc}A_{\mu
}^{c}\right) c^{b}\right) \;,  \label{eq1}
\end{equation}
where $\overline{c}^{a}$, $c^{a}$ stand for the Faddeev-Popov ghosts. The
Lagrange multiplier $b^{a}$ has been introduced to implement the gauge
condition
\begin{equation}
\partial _{\mu }A_{\mu }^{a}=-\alpha b^{a}\;,  \label{eqm}
\end{equation}
from which it follows
\begin{equation}
\int d^{4}x\left( b^{a}\partial A^{a}+\frac{\alpha }{2}b^{a}b^{a}\right)
\Rightarrow -\frac{1}{2\alpha }\int d^{4}x\;\left( \partial _{\mu }A_{\mu
}^{a}\right) ^{2}\;.  \label{eb}
\end{equation}
From the relation $\left( \mathrm{{\ref{eqm}}}\right) $ we see
that the field $A_{\mu }^{a}$ is not transverse, due to the
presence of the gauge parameter $\alpha $. Of course, in the limit
$\alpha \rightarrow 0$, we recover the Landau gauge, $\partial
_{\mu }A_{\mu }^{a}=0$. In what follows, it will be useful to
decompose the gauge field $A_{\mu }^{a}$ into transverse and
longitudinal components, according to
\begin{equation}
A_{\mu }^{a}=A_{\mu }^{aT}+A_{\mu }^{aL}\;,  \label{ad}
\end{equation}
with
\begin{eqnarray}
A_{\mu }^{aT} &=&\left( \delta _{\mu \nu }-\frac{\partial _{\mu }\partial
_{\nu }}{\partial ^{2}}\right) A_{\nu }^{a}\;,  \nonumber \\
\partial _{\mu }A_{\mu }^{aT} &=&0\;,  \label{atr}
\end{eqnarray}
and
\begin{eqnarray}
A_{\mu }^{aL} &=&\frac{\partial _{\mu }\partial _{\nu }}{\partial ^{2}}%
A_{\nu }^{a}=-\alpha \frac{\partial _{\mu }}{\partial ^{2}}b^{a}\;,
\nonumber \\
\partial _{\mu }A_{\mu }^{aL} &=&-\alpha b^{a}\;.  \label{al}
\end{eqnarray}
As already underlined, we shall restrict ourselves to the case in which $%
\alpha $ is small, \textit{i.e. }$\alpha \ll 1$, so that the longitudinal
component $A_{\mu }^{aL}$ in eq.$\left( \mathrm{{\ref{ad}}}\right) $ can be
considered as a small perturbation with respect to the transverse part $%
A_{\mu }^{aT}$. \newline
\newline
Let us proceed by proving the following statement.

\begin{itemize}
\item  \textbf{Statement}

If the transverse component $A_{\mu }^{aT}$ of the gauge field $A_{\mu
}^{a}=\left( A_{\mu }^{aT}+A_{\mu }^{aL}\right) $ belongs to the Gribov
region $\Omega $, \textit{i.e.} $A_{\mu }^{aT}\in \,\Omega $, $\Omega
=\left\{ A_{\mu }^{aT},\;\partial _{\mu }A_{\mu }^{aT}=0,\mathcal{\;}%
-\partial _{\mu }\left( \partial _{\mu }\delta ^{ab}-gf^{abc}A_{\mu
}^{cT}\right) >0\right\} $, then the Faddeev-Popov operator $\mathcal{M}%
^{ab}(A)=-\partial _{\mu }\left( \partial _{\mu }\delta ^{ab}-gf^{abc}A_{\mu
}^{c}\right) $ has no zero modes.

\textbf{Proof}

The proof of this statement is done by assuming the converse.
Suppose indeed that the operator $\mathcal{M}^{ab}(A)$ has a zero
mode, \textit{i.e. }it exists a $\widetilde{\varphi }^{a}(x,\alpha
)$ such that
\begin{equation}
-\partial _{\mu }\left( \partial _{\mu }\delta ^{ab}-gf^{abc}A_{\mu
}^{cT}-gf^{abc}A_{\mu }^{cL}\right) \widetilde{\varphi }^{b}(x,\alpha )=0\;,
\label{zm1}
\end{equation}
which, from eq.$\left( \mathrm{{\ref{al}}}\right) $, becomes
\begin{equation}
-\partial _{\mu }\left( \partial _{\mu }\delta ^{ab}-gf^{abc}A_{\mu
}^{cT}+\alpha gf^{abc}\left( \frac{\partial _{\mu }}{\partial ^{2}}%
b^{c}\right) \right) \widetilde{\varphi }^{b}(x,\alpha )=0\;.  \label{zm2}
\end{equation}
Since $\alpha $ is small we perform a Taylor expansion of $\widetilde{%
\varphi }^{a}(x,\alpha )$, namely
\begin{equation}
\widetilde{\varphi }^{a}(x,\alpha )=\sum_{n=0}^{\infty }\alpha ^{n}%
\widetilde{\varphi }_{n}^{a}(x)\;,  \label{zm3}
\end{equation}
with
\begin{equation}
\widetilde{\varphi }_{0}^{a}(x)=\left. \widetilde{\varphi }^{a}(x,\alpha
)\right| _{\alpha =0}\;.  \label{zm4}
\end{equation}
Equation $\left( \mathrm{{\ref{zm2}}}\right) $ splits thus in a tower of
equations, given by
\begin{equation}
-\partial _{\mu }\left( \partial _{\mu }\delta ^{ab}-gf^{abc}A_{\mu
}^{cT}\right) \widetilde{\varphi }_{0}^{b}(x)=0\;,  \label{zm5}
\end{equation}
\begin{equation}
-\partial _{\mu }\left( \partial _{\mu }\delta ^{ab}-gf^{abc}A_{\mu
}^{cT}\right) \widetilde{\varphi }_{1}^{b}(x)-\partial _{\mu }\left(
gf^{abc}\left( \frac{\partial _{\mu }}{\partial ^{2}}b^{c}\right) \widetilde{%
\varphi }_{0}^{b}(x)\right) =0\;,  \label{zm6}
\end{equation}
\begin{equation}
-\partial _{\mu }\left( \partial _{\mu }\delta ^{ab}-gf^{abc}A_{\mu
}^{cT}\right) \widetilde{\varphi }_{2}^{b}(x)-\partial _{\mu }\left(
gf^{abc}\left( \frac{\partial _{\mu }}{\partial ^{2}}b^{c}\right) \widetilde{%
\varphi }_{1}^{b}(x)\right) =0\;,  \label{zm7}
\end{equation}
\begin{equation}
-\partial _{\mu }\left( \partial _{\mu }\delta ^{ab}-gf^{abc}A_{\mu
}^{cT}\right) \widetilde{\varphi }_{3}^{b}(x)-\partial _{\mu }\left(
gf^{abc}\left( \frac{\partial _{\mu }}{\partial ^{2}}b^{c}\right) \widetilde{%
\varphi }_{2}^{b}(x)\right) =0\;,  \label{zm8}
\end{equation}
and so on. \\\\Moreover, since $A_{\mu }^{aT}$ belongs to the Gribov region $\Omega $, $%
A_{\mu }^{aT}\in \,\Omega $, it follows that $\widetilde{\varphi }_{0}^{b}(x)
$ in the first equation $\left( \mathrm{{\ref{zm5}}}\right) $ necessarily
vanishes, \textit{i.e. }$\widetilde{\varphi }_{0}^{b}(x)=0$, since the
operator $-\partial _{\mu }\left( \partial _{\mu }\delta
^{ab}-gf^{abc}A_{\mu }^{cT}\right) $ has no zero modes in $\Omega $.
Furthermore, setting $\widetilde{\varphi }_{0}^{b}(x)=0$ in the second
equation $\left( \mathrm{{\ref{zm6}}}\right) $, we get
\begin{equation}
-\partial _{\mu }\left( \partial _{\mu }\delta ^{ab}-gf^{abc}A_{\mu
}^{cT}\right) \widetilde{\varphi }_{1}^{b}(x)=0\;,  \label{zm9}
\end{equation}
which, in turn, implies the vanishing of $\widetilde{\varphi }_{1}^{b}(x)$,
\textit{i.e. }$\widetilde{\varphi }_{1}^{b}(x)=0$. As a consequence, eq.$%
\left( \mathrm{{\ref{zm7}}}\right) $ reads
\begin{equation}
-\partial _{\mu }\left( \partial _{\mu }\delta ^{ab}-gf^{abc}A_{\mu
}^{cT}\right) \widetilde{\varphi }_{2}^{b}(x)=0\;,  \label{zm10}
\end{equation}
from which $\widetilde{\varphi }_{2}^{b}(x)=0$. It is apparent thus that the
argument can be easily iterated to the whole tower of equations, implying
that $\widetilde{\varphi }^{b}(x,\alpha )$ vanishes, $\widetilde{\varphi }%
^{b}(x,\alpha )=0$. This concludes the proof of the statement.
\end{itemize}

\noindent In particular, from this result it follows that if
$A_{\mu }^{aT}$
belongs to the Gribov region, $A_{\mu }^{aT}\in \,\Omega $, the field $%
A_{\mu }^{a}=\left( A_{\mu }^{aT}+A_{\mu }^{aL}\right) $ does not possess
Gribov copies which are closely related, \textit{i.e.} obtained from $A_{\mu
}^{a}$ through an infinitesimal gauge transformation
\begin{equation}
\widetilde{A}_{\mu }^{a}=A_{\mu }^{a}+\left( \partial _{\mu }\delta
^{ab}-gf^{abc}A_{\mu }^{c}\right) \omega ^{b}\;,  \label{zm11}
\end{equation}
where $\omega ^{a}(x)$ denotes the parameter of the gauge transformation.
Indeed, from the condition
\begin{equation}
\partial _{\mu }\widetilde{A}_{\mu }^{a}=\partial _{\mu }A_{\mu }^{a}\;,
\label{zm12}
\end{equation}
we should have
\begin{equation}
\partial _{\mu }\left( \partial _{\mu }\delta ^{ab}-gf^{abc}A_{\mu
}^{c}\right) \omega ^{b}\;=0\;,  \label{zm13}
\end{equation}
which, however, has no solution for $\omega ^{a}(x)$.

\subsection{Restriction of the domain of integration in the path-integral}

The results obtained in the previous sections suggest to restrict the domain
of integration in the path integral to the region $\widetilde{\Omega }$
defined as
\begin{equation}
\widetilde{\Omega }\equiv \left\{ A_{\mu }^{a};\;A_{\mu }^{a}=\;A_{\mu
}^{aT}+A_{\mu }^{aL},\;A_{\mu }^{aT}\in \,\Omega \right\} \;,  \label{r1}
\end{equation}
\textit{i.e.} $\widetilde{\Omega }$ is the set of all connections whose
transverse part $A_{\mu }^{aT}$ belongs to the Gribov region $\Omega
=\left\{ A_{\mu }^{aT},\;\partial _{\mu }A_{\mu }^{aT}=0,\mathcal{\;}%
-\partial _{\mu }\left( \partial _{\mu }\delta ^{ab}-gf^{abc}A_{\mu
}^{cT}\right) >0\right\} $. This choice is motivated by the following
considerations:

\begin{itemize}
\item  Since the gauge parameter $\alpha $ is small, $\alpha \ll 1$, the
region $\widetilde{\Omega }$ can be regarded as a deformation of the Gribov
region $\Omega $. It is apparent in fact that in the limit $\alpha
\rightarrow 0$ the region $\Omega $ is recovered, namely

\begin{equation}
\lim_{\alpha \rightarrow 0}\widetilde{\Omega }=\Omega \;.  \label{r2}
\end{equation}

\item  As discussed before, the Faddeev-Popov operator, $\mathcal{M}%
^{ab}(A)=-\partial _{\mu }\left( \partial _{\mu }\delta
^{ab}-gf^{abc}A_{\mu }^{c}\right) $, has no zero modes within
$\widetilde{\Omega }$. Therefore, the restriction to
$\widetilde{\Omega }$ allows us to get rid of the Gribov copies
which can be obtained through infinitesimal gauge transformations.

\item  The effective implementation in the path-integral of the
restriction of the domain of integration to the region
$\widetilde{\Omega }$ can be done by repeating the no-pole
prescription of Gribov's original work \cite{Gribov:1977wm}.
Indeed,
since the transverse component $A_{\mu }^{aT}$ of any field belonging to $%
\widetilde{\Omega }$ lies in the Gribov region $\Omega $, to implement the
restriction to $\widetilde{\Omega }$ it will be sufficient to require that
the Green function $\mathcal{G}_{tr}(k)$ of eq.$\left( \mathrm{{\ref{gtr}}}%
\right) $ has no poles for any $A_{\mu }^{aT}\in \,\Omega $, except for a
singularity at $k^{2}=0$, corresponding to the Gribov horizon $\partial
\,\Omega $.
\end{itemize}

\noindent Thus, for the partition function of Yang-Mills theories in linear
covariant gauges, we write
\begin{eqnarray}
\mathcal{Z} &=&\int DADb\delta (\partial A+\alpha b)\det \left( \mathcal{M}%
^{ab}(A)\right) e^{-\left( \frac{1}{4}\int d^{4}xF_{\mu \nu }^{a}F_{\mu \nu
}^{a}-\int d^{4}x\left( b^{a}\partial A^{a}+\frac{\alpha }{2}%
b^{a}b^{a}\right) \right) }\mathcal{V}(\widetilde{\Omega })\;  \nonumber \\
&=&\int DA\;\det \left( -\partial _{\mu }\left( \delta ^{ab}\partial _{\mu
}-gf^{abc}A_{\mu }^{c}\right) \right) e^{-\left( \frac{1}{4}\int
d^{4}xF_{\mu \nu }^{a}F_{\mu \nu }^{a}+\frac{1}{2\alpha }\int d^{4}x\left(
\partial A^{a}\right) ^{2}\right) }\mathcal{V}(\widetilde{\Omega })\;,\;
\label{r3}
\end{eqnarray}
where the factor $\mathcal{V}(\widetilde{\Omega })$ implements the
restriction to the region $\widetilde{\Omega }$. An explicit
characterization of $\mathcal{V}(\widetilde{\Omega })$ and of its
consequences on the infrared behavior of the gluon propagator and of $%
\mathcal{G}_{tr}(k)$ will be discussed in the next section.
\\\\Finally, it is worth to spend a few words on the aspects which
remain uncovered by the present investigation. Even if the
restriction to the region $\widetilde{\Omega }$ allows us to get
rid of the Gribov copies which are closely related, {\it i.e.}
attainable by infinitesimal gauge transformations, we still lack a
treatment of the copies which cannot be attained by infinitesimal
transformations. This task is beyond our present possibilities, as
the knowledge of a suitable auxiliary functional associated to the
linear covariant gauges would be required. Nevertheless, since we
are limiting ourselves to small values of $\alpha $, the
restriction to the region $\widetilde{\Omega }$ looks quite
natural.

\subsection{Characterization of the factor $\mathcal{V}(\widetilde{\Omega })$}

As already remarked, the factor $\mathcal{V}(\widetilde{\Omega })$ can be
accommodated for by requiring that the Green function $\mathcal{G}_{tr}(k)$
of eq.$\left( \mathrm{{\ref{gtr}}}\right) $ has no poles for a given
nonvanishing value of the momentum $k$. This condition can be understood by
recalling that the region $\Omega $ is defined as the set of all transverse
gauge connections $\left\{ A_{\mu }^{Ta}\right\} $, $\partial _{\mu }A_{\mu
}^{Ta}=0,$ for which the operator $\mathcal{M}^{ab}(A^{T})$ is positive
definite, \textit{i.e.\thinspace }$\mathcal{M}^{ab}(A^{T})=-\partial _{\mu
}\left( \partial _{\mu }\delta ^{ab}-gf^{abc}A_{\mu }^{Tc}\right) >0$. This
implies that the inverse operator $\left[ -\partial _{\mu }\left( \partial
_{\mu }\delta ^{ab}-gf^{abc}A_{\mu }^{Tc}\right) \right] ^{-1}$, and thus $%
\mathcal{G}_{tr}(k)$, can become large only when approaching the horizon $%
\partial \Omega $, which corresponds in fact to $k=0$ \cite{Gribov:1977wm}. The
quantity $\mathcal{G}_{tr}(k)$ can be evaluated order by order in
perturbation theory. Repeating the same calculation of
\cite{Gribov:1977wm}, we find that, up to the second order
\begin{equation}
\mathcal{G}_{tr}(k)\approx \frac{1}{k^{2}}\frac{1}{1-\rho (k,A^{T})}\;,
\label{g1}
\end{equation}
with $\rho (k,A^{T})$ given by
\begin{eqnarray}
\rho (k,A^{T}) &=&\frac{N}{N^{2}-1}\frac{g^{2}}{V}\frac{1}{k^{2}}\sum_{q}%
\frac{k_{\mu }(k-q)_{\nu }}{\left( k-q\right) ^{2}}A_{\mu }^{Ta}(q)A_{\nu
}^{Ta}(-q)\;  \nonumber \\
&=&\frac{N}{N^{2}-1}\frac{g^{2}}{V}\frac{k_{\mu }k_{\nu }}{k^{2}}\sum_{q}%
\frac{1}{\left( k-q\right) ^{2}}A_{\mu }^{Ta}(q)A_{\nu }^{Ta}(-q)\;\;,
\label{g2}
\end{eqnarray}
$V$ being the Euclidean space-time volume. According to
\cite{Gribov:1977wm}, the no-pole condition for
$\mathcal{G}_{tr}(k)$ reads
\begin{eqnarray}
\rho (0,A^{T}) &<&1\;,  \nonumber \\
\rho (0,A^{T}) &=&\frac{1}{4}\frac{N}{N^{2}-1}\frac{g^{2}}{V}\sum_{q}\frac{1%
}{q^{2}}\left( A_{\lambda }^{Ta}(q)A_{\lambda }^{Ta}(-q)\right) \;.
\label{g3}
\end{eqnarray}
Therefore, for the factor $\mathcal{V}(\widetilde{\Omega })$ in eq.$\left(
\ref{r3}\right) $ we have
\begin{equation}
\mathcal{V}(\widetilde{\Omega })=\theta (1-\rho (0,A^{T}))\;,  \label{g4}
\end{equation}
where $\theta (x)$ stands for the step function\footnote{$\theta (x)=1$ if $%
x>0$, $\theta (x)=0$ if $x<0$.}.\

\section{The gluon propagator}

In order to study the gluon propagator, it is sufficient to retain
only the quadratic terms in expression $\left( \ref{r3}\right) $
which contribute to the two-point correlation function
$\left\langle A_{\mu }^{a}(k)A_{\nu }^{b}(-k)\right\rangle $.
Making use of the integral representation for the step function
\begin{equation}
\theta (1-\rho (0,A^{T}))=\int_{-i\infty +\varepsilon }^{i\infty
+\varepsilon }\frac{d\eta }{2\pi i\eta }e^{\eta (1-\rho (0,A^{T}))}\;,
\label{p1}
\end{equation}
for the partition function $\left( \ref{r3}\right) $ one get
\begin{equation}
\mathcal{Z}_{\mathrm{quadr}}=\mathcal{N}\int DA\frac{d\eta }{2\pi i}e^{\eta
-\log \eta }e^{-\left( S_{\mathrm{quadr}}+\eta \rho (0,A^{T})\right) }\;,
\label{p2}
\end{equation}
where $\mathcal{N}$ is a constant factor and $S_{\mathrm{quadr}}$ stands for
the quadratic part of the quantized Yang-Mills action
\begin{equation}
S_{\mathrm{quadr}}=\frac{1}{2}\sum_{q}\left( q^{2}A_{\mu }^{a}(q)A_{\mu
}^{a}(-q)-\left( 1-\frac{1}{\alpha }\right) q_{\mu }q_{\nu }A_{\mu
}^{a}(q)A_{\nu }^{a}(-q)\right) \;.  \label{p3}
\end{equation}
From
\begin{eqnarray}
A_{\mu }^{Ta}(q)A_{\mu }^{Ta}(-q) &=&\left( \delta _{\mu \nu }-\frac{q_{\mu
}q_{\nu }}{q^{2}}\right) \left( \delta _{\mu \sigma }-\frac{q_{\mu
}q_{\sigma }}{q^{2}}\right) A_{\nu }^{a}(q)A_{\sigma }^{a}(-q)  \nonumber \\
&=&A_{\mu }^{a}(q)A_{\mu }^{a}(-q)-\frac{q_{\mu }q_{\nu }}{q^{2}}A_{\mu
}^{a}(q)A_{\nu }^{a}(-q)\;,  \label{p4}
\end{eqnarray}
it follows that
\begin{equation}
\mathcal{Z}_{\mathrm{quadr}}=\mathcal{N}\int DA\frac{d\eta }{2\pi i}e^{\eta
-\log \eta }e^{-\frac{1}{2}\sum_{q}A_{\mu }^{a}(q)\mathcal{Q}_{\mu \nu
}^{ab}A_{\nu }^{b}(-q)}\;,  \label{p5}
\end{equation}
with
\begin{equation}
\mathcal{Q}_{\mu \nu }^{ab}=\left( \left( q^{2}+\frac{\eta Ng^{2}}{N^{2}-1}%
\frac{1}{2V}\frac{1}{q^{2}}\right) \delta _{\mu \nu }-q_{\mu }q_{\nu }\left(
\left( 1-\frac{1}{\alpha }\right) +\frac{\eta Ng^{2}}{N^{2}-1}\frac{1}{2V}%
\frac{1}{q^{4}}\right) \right) \delta ^{ab}\;.  \label{p6}
\end{equation}
Integrating over the gauge field, one has
\begin{equation}
\mathcal{Z}_{\mathrm{quadr}}=\mathcal{N}\int \frac{d\eta }{2\pi i}e^{\eta
-\log \eta }\left( \det \mathcal{Q}_{\mu \nu }^{ab}\right) ^{-\frac{1}{2}}=%
\mathcal{N}^{\prime }\int \frac{d\eta }{2\pi i}e^{f(\eta )}\;,  \label{p7}
\end{equation}
where $f(\eta )$ is given by
\begin{equation}
f(\eta )=\eta -\log \eta -\frac{3}{2}(N^{2}-1)\sum_{q}\log \left( q^{4}+%
\frac{\eta Ng^{2}}{N^{2}-1}\frac{1}{2V}\right) \;.  \label{p8}
\end{equation}
Following \cite{Gribov:1977wm}, the expression $\left(
\mathrm{{\ref{p7}}}\right) $ can be now evaluated at the saddle
point, namely
\begin{equation}
\mathcal{Z}_{\mathrm{quadr}}\approx e^{f(\eta _{0})}\;,  \label{p9}
\end{equation}
where $\eta _{0}$ is determined by the minimum condition
\begin{equation}
1-\frac{1}{\eta _{0}}-\frac{3}{4}\frac{Ng^{2}}{V}\sum_{q}\frac{1}{\left(
q^{4}+\frac{\eta _{0}Ng^{2}}{N^{2}-1}\frac{1}{2V}\right) }=0\;.  \label{p10}
\end{equation}
Taking the thermodynamic limit, $V\rightarrow \infty $, and
introducing the Gribov parameter $\gamma $ \cite{Gribov:1977wm}
\begin{equation}
\gamma ^{4}=\frac{\eta _{0}Ng^{2}}{N^{2}-1}\frac{1}{2V}\;\;,\;\;\;\;V%
\rightarrow \infty \;,  \label{p11}
\end{equation}
we get the gap equation
\begin{equation}
\frac{3}{4}Ng^{2}\int \frac{d^{4}q}{\left( 2\pi \right) ^{4}}\frac{1}{%
q^{4}+\gamma ^{4}}=1\;,  \label{p12}
\end{equation}
where the term $1/\eta _{0}$ in $\left( \mathrm{{\ref{p10}}}\right) $ has
been neglected in the thermodynamic limit. To obtain the gauge propagator,
we can now go back to the expression for $\mathcal{Z}_{\mathrm{quadr}}$
which, after substituting the saddle point value $\eta =\eta _{0}$, becomes
\begin{equation}
\mathcal{Z}_{\mathrm{quadr}}=\mathcal{N}\int DAe^{-\frac{1}{2}\sum_{q}A_{\mu
}^{a}(q)\mathcal{Q}_{\mu \nu }^{ab}A_{\nu }^{b}(-q)}\;,  \label{p13}
\end{equation}
with
\begin{eqnarray}
\mathcal{Q}_{\mu \nu }^{ab} &=&\left( \left( q^{2}+\frac{\gamma ^{4}}{q^{2}}%
\right) \delta _{\mu \nu }-q_{\mu }q_{\nu }\left( \left( 1-\frac{1}{\alpha }%
\right) +\frac{\gamma ^{4}}{q^{4}}\right) \right) \delta ^{ab}\;  \nonumber
\\
&=&\left( \left( q^{2}+\frac{\gamma ^{4}}{q^{2}}\right) \left( \delta _{\mu
\nu }-\frac{q_{\mu }q_{\nu }}{q^{2}}\right) +\frac{q_{\mu }q_{\nu }}{q^{2}}%
\left( \frac{q^{2}}{\alpha }\right) \right) \delta ^{ab}\;.  \label{p14}
\end{eqnarray}
Evaluating the inverse of $\mathcal{Q}_{\mu \nu }^{ab}$ , we get the gluon
propagator
\begin{equation}
\left\langle A_{\mu }^{a}(q)A_{\nu }^{b}(-q)\right\rangle =\delta
^{ab}\left( \frac{q^{2}}{q^{4}+\gamma ^{4}}\left( \delta _{\mu \nu }-\frac{%
q_{\mu }q_{\nu }}{q^{2}}\right) \;+\frac{\alpha }{q^{2}}\frac{q_{\mu }q_{\nu
}}{q^{2}}\right) \;.  \label{p15}
\end{equation}
A few remarks are now in order. \\\\Let us first observe that the gap equation $\left( \mathrm{{\ref{p12}}}%
\right) $ defining the parameter $\gamma $ has the same form of
that obtained by Gribov in the Landau gauge \cite{Gribov:1977wm}.
This is an expected result, since the factor $\rho (0,A^{T})$ in
equation $\left( \mathrm{{\ref{g3}}}\right) $ depends only on the
transverse component $A_{\mu }^{Ta}$. \\\\As it is apparent from
the expression $\left( \mathrm{{\ref{p15}}}\right) $, the
transverse component of the gluon propagator is suppressed in the
infrared, while the longitudinal component is left unchanged.
Moreover, as we shall see later, the behavior of the longitudinal
part turns out to be modified once the gluon condensate
$\left\langle A_{\mu }^{a}A_{\mu }^{a}\right\rangle $ is taken
into account.
\\\\Finally, in the limit $\alpha \rightarrow 0$, the results of
the Landau gauge are recovered.

\subsection{The infrared behavior of $\mathcal{G}_{tr}(k)$}

Let us discuss now the infrared behavior of the Green function $\mathcal{G}%
_{tr}(k)$ of eq.$\left( \mathrm{{\ref{gtr}}}\right) $, which is obtained
upon contraction of the gauge fields in the expression $\left( \mathrm{{\ref
{g2}}}\right) $, namely
\begin{equation}
\mathcal{G}_{tr}(k)\approx \frac{1}{k^{2}}\frac{1}{1-\rho (k)}\;,  \label{i1}
\end{equation}
with
\begin{equation}
\rho (k)=g^{2}\frac{N}{N^{2}-1}\frac{k_{\mu }k_{\nu }}{k^{2}}\int \frac{%
d^{4}q}{\left( 2\pi \right) ^{4}}\frac{1}{\left( k-q\right) ^{2}}%
\left\langle A_{\mu }^{Ta}(q)A_{\nu }^{Ta}(-q)\right\rangle \;.  \label{i2}
\end{equation}
From the expression of the gluon propagator in eq.$\left( \mathrm{{\ref{p15}}%
}\right) $, it follows
\begin{equation}
\rho (k)=g^{2}N\frac{k_{\mu }k_{\nu }}{k^{2}}\int \frac{d^{4}q}{\left( 2\pi
\right) ^{4}}\frac{1}{\left( k-q\right) ^{2}}\frac{q^{2}}{q^{4}+\gamma ^{4}}%
\left( \delta _{\mu \nu }-\frac{q_{\mu }q_{\nu }}{q^{2}}\right) \;.
\label{i3}
\end{equation}
Furthermore, from the gap equation $\left( \mathrm{{\ref{p12}}}\right) $, it
turns out
\begin{equation}
Ng^{2}\int \frac{d^{4}q}{\left( 2\pi \right) ^{4}}\frac{1}{q^{4}+\gamma ^{4}}%
\left( \delta _{\mu \nu }-\frac{q_{\mu }q_{\nu }}{q^{2}}\right) =\delta
_{\mu \nu }\;,  \label{i4}
\end{equation}
so that
\begin{equation}
1-\rho (k)=Ng^{2}\frac{k_{\mu }k_{\nu }}{k^{2}}\int \frac{d^{4}q}{\left(
2\pi \right) ^{4}}\frac{k^{2}-2qk}{\left( k-q\right) ^{2}}\frac{1}{%
q^{4}+\gamma ^{4}}\left( \delta _{\mu \nu }-\frac{q_{\mu }q_{\nu }}{q^{2}}%
\right) \;.  \label{i5}
\end{equation}
Note that the integral in the right hand side of eq.$\left(
\mathrm{{\ref
{i5}}}\right) $ is convergent and nonsingular at $k=0$. Therefore, for $%
k\approx 0$,
\begin{equation}
\left( 1-\rho (k)\right) _{k\approx 0}\approx \frac{3Ng^{2}\mathcal{I}}{4}%
k^{2}\;,  \label{i6}
\end{equation}
where $\mathcal{I}$ stands for the value of the integral
\begin{equation}
\mathcal{I=}\int \frac{d^{4}q}{\left( 2\pi \right) ^{4}}\frac{1}{%
q^{2}(q^{4}+\gamma ^{4})} = \frac{1}{32\pi} \frac{1}{\gamma^2}\;.
\label{i7}
\end{equation}
Finally, for the infrared behavior of the Green function $\mathcal{G}_{tr}(k)
$ we get
\begin{equation}
\left( \mathcal{G}_{tr}\right) _{k\approx 0}\approx \frac{4}{3Ng^{2}\mathcal{%
I}}\frac{1}{k^{4}}\;.  \label{i8}
\end{equation}
One sees thus that $\mathcal{G}_{tr}(k)$ is enhanced in the infrared,
obeying the Gribov-Zwanziger condition $\lim_{k\rightarrow 0}\left( k^{2}%
\mathcal{G}_{tr}(k)\right) ^{-1}=0$.
\subsection{Analysis of the ghost propagator}
For the sake of completeness, let us discuss here the infrared
behavior of the ghost two-point function, $\mathcal{G}_{gh}(k)$,
given by
\begin{equation}
\mathcal{G}_{gh}(k) =\frac{1}{N^{2}-1}\sum_{ab} \delta ^{ab}\left\langle \overline{c}%
^{a}(k)c^{b}(-k)\right\rangle \approx
\frac{1}{k^2}\frac{1}{1-\omega(k)} \;,  \label{gh1}
\end{equation}
with
\begin{equation}
\omega(k)=\frac{N}{N^{2}-1}\frac{g^{2}}{k^{2}}\int \frac{%
d^{4}q}{\left( 2\pi \right) ^{4}}
\frac{k_\mu(k-q)_\nu}{(k-q)^2}\left\langle A_{\mu }^{a}(q)A_{\nu
}^{a}(-q)\right\rangle \;. \label{gh2}
\end{equation}
Making use of the expression for the gluon propagator in
eq.$\left( \mathrm{{\ref{p15}}}\right)$ and of the equation
$\left( \mathrm{{\ref{i4}}}\right)$, it follows that, in the
infrared,
\begin{equation}
1-\omega(k) \approx \frac{3Ng^2}{128\pi}\frac{k^2}{\gamma^2}
-\frac{\alpha Ng^2}{k^2} \int \frac{d^{4}q}{\left( 2\pi \right)
^{4}} \frac{k_\mu(k-q)_\nu}{(k-q)^2}\frac{q_\mu q_\nu}{q^4} \;.
\label{gh3}
\end{equation}
The second term in the right hand-side of eq.$\left(
\mathrm{{\ref{gh3}}}\right)$ can be easily evaluated by means of
the dimensional regularization. Adopting the $\overline{MS}$
scheme, the final expression for the factor $(1-\omega(k))$ is
found
\begin{equation}
(1-\omega(k)) \approx \frac{3Ng^2}{128\pi}\frac{k^2}{\gamma^2}
-\frac{\alpha Ng^2}{64\pi^2}\;log{\frac{k^2}{{\overline\mu}^2}}
\;. \label{gh4}
\end{equation}
One sees that, in the present case, the Gribov-Zwanziger horizon
condition is jeopardized by the term
$log{\frac{k^2}{{\overline\mu}^2}}$ in expression $\left(
\mathrm{{\ref{gh4}}}\right)$. As it is apparent from the presence
of the gauge parameter $\alpha$, this term is due to the
contribution of the longitudinal components of the gauge field.
Note that the longitudinal modes do not contribute to the Green
function $\mathcal{G}_{tr}(k)$ in eq.$\left(
\mathrm{{\ref{i1}}}\right)$.

\section{Inclusion of the dimension two condensate $\left\langle A_{\mu
}^{a}A_{\mu }^{a}\right\rangle $}

The dimension two gluon condensate $\left\langle A_{\mu
}^{a}A_{\mu }^{a}\right\rangle $ has received much attention in
the last years
\cite{Gubarev:2000nz,Gubarev:2000eu,Verschelde:2001ia,Kondo:2001tm,Kondo:2001nq,Boucaud:2001st,Boucaud:2000nd,
Dudal:2003vv,Dudal:2002pq,Browne:2003uv,RuizArriola:2004en}. This
condensate turns out to contribute to the gluon two-point
function, as observed in \cite{Lavelle:1988eg} within the operator
product expansion. As such, it has to be taken into account when
discussing the gluon propagator. A renormalizable effective
potential for $\left\langle A_{\mu }^{a}A_{\mu }^{a}\right\rangle
$ in linear covariant gauges has been constructed and evaluated in
analytic form in \cite{Dudal:2003by}. The output of this
investigation is that a nonvanishing value of the condensate
$\left\langle A_{\mu }^{a}A_{\mu }^{a}\right\rangle $ is favoured
since it lowers the vacuum energy. As a consequence, a dynamical
tree level gluon mass is generated in
the gauge fixed Lagrangian \cite{Dudal:2003by}. The inclusion of the condensate $%
\left\langle A_{\mu }^{a}A_{\mu }^{a}\right\rangle $ in the
present framework can be done along the lines outlined in
\cite{Sobreiro:2004us}, where the effects of the Gribov copies on
the gluon and ghost propagators in the presence of $\left\langle
A_{\mu }^{a}A_{\mu }^{a}\right\rangle $ have been worked out in
the Landau gauge. Let us begin by giving a brief account of the
dynamical mass generation in linear covariant gauges. Following
\cite{Dudal:2003by}, the dynamical mass generation in these gauges
is described by the following action
\begin{equation}
S(A,\sigma )=S_{YM}+S_{gf}+S_{\sigma }\;,  \label{m1}
\end{equation}
where $S_{YM}$, $S_{gf}$ are the Yang-Mills and the gauge fixing
terms, as given in eq.$\left( \ref{eq1}\right) $. The term
$S_{\sigma }$ in eq.$\left( \ref{m1}\right) $ contains the
auxiliary scalar field $\sigma $ and reads
\begin{equation}
S_{\sigma }=\int d^{4}x\left( \frac{\sigma ^{2}}{2g^{2}\zeta }+\frac{1}{2}%
\frac{\sigma }{g\zeta }A_{\mu }^{a}A_{\mu }^{a}+\frac{1}{8\zeta }\left(
A_{\mu }^{a}A_{\mu }^{a}\right) ^{2}\;\right) .  \label{m2}
\end{equation}
The introduction of the auxiliary field $\sigma $ allows us to
study the condensation of the local operator $A_{\mu }^{a}A_{\mu
}^{a}$. In fact, as shown in \cite{Dudal:2003by}, the following
relation holds
\begin{equation}
\left\langle \sigma \right\rangle =-\frac{g}{2}\left\langle A_{\mu
}^{a}A_{\mu }^{a}\right\rangle \;.  \label{m3}
\end{equation}
The dimensionless parameter $\zeta $ in expression $\left(
\ref{m2}\right) $ is needed to account for the ultraviolet
divergences present in the vacuum correlation function
$\left\langle A^{2}(x)A^{2}(y)\right\rangle $. For the details of
the renormalizability properties of the local operator $A_{\mu
}^{a}A_{\mu }^{a}$ in linear covariant gauges we refer to
\cite{Dudal:2003np}. The action $S(A,\sigma )$ is the starting
point for evaluating the
renormalizable effective potential $V(\sigma )$ for the auxiliary field $%
\sigma $, obeying the renormalization group equations. The minimum
of $V(\sigma )$ occurs for a nonvanishing vacuum expectation value
of the auxiliary field \cite{Dudal:2003by}, \textit{i.e.
}$\left\langle
\sigma \right\rangle \neq 0$. In particular, from expression $\left( \ref{m1}%
\right) $, the first order induced dynamical gluon mass is found to be
\begin{equation}
m^{2}=\frac{g\left\langle \sigma \right\rangle }{\zeta _{0}}\;,  \label{m4}
\end{equation}
where $\zeta _{0}$ is the first contribution to the parameter
$\zeta $, given by \cite{Dudal:2003by}
\begin{eqnarray}
\zeta  &=&\frac{\zeta _{0}}{g^{2}}+\zeta _{1}+O(g^{2})\;,  \nonumber \\
\zeta _{0} &=&\frac{3\left( 78-26\alpha ^{2}+3\alpha ^{3}+18\alpha \log
\alpha \right) }{2\left( 3\alpha -13\right) ^{2}}\frac{\left( N^{2}-1\right)
}{N}\;.  \label{m5}
\end{eqnarray}
We remind here that in linear covariant gauges, the Faddeev-Popov ghosts $%
\overline{c}^{a},$ $c^{a}$ remain massless, due to the absence of mixing
between the composite operators $A_{\mu }^{a}(x)A_{\mu }^{a}(x)$ and $\overline{c}%
^{a}(x)c^{a}(x)$. This stems from additional Ward identities
present in these
gauges \cite{Dudal:2003np}, which forbid the appearance of the term $\overline{c}%
^{a}(x)c^{a}(x)$.

\subsection{Infrared behavior of the gluon propagator in the presence of $%
\left\langle A_{\mu }^{a}A_{\mu }^{a}\right\rangle $}

It is worth underlining that the action $S(A,\sigma )$ leads to a
partition function which is still plagued by the Gribov copies. It
might be useful to note in fact that the action $\left(
S_{YM}+S_{\sigma }\right) $ is left invariant by the local gauge
transformations
\begin{eqnarray}
\delta A_{\mu }^{a} &=&-D_{\mu }^{ab}\omega ^{b}\;,\;\;\;\;\;\;  \label{m6}
\\
\delta \sigma  &=&gA_{\mu }^{a}\partial _{\mu }\omega ^{a}\;,\;  \nonumber
\end{eqnarray}
\begin{equation}
\delta \left( S_{YM}+S_{\sigma }\right) =0\;.  \label{m7}
\end{equation}
Therefore, implementing the restriction to the region $\widetilde{\Omega }$,
for the partition function we obtain
\begin{equation}
\mathcal{Z}=\int DA\;D\sigma \det \left( -\partial _{\mu }\left( \delta
^{ab}\partial _{\mu }+gf^{abc}A_{\mu }^{c}\right) \right) e^{-\left( \frac{1%
}{4}\int d^{4}xF_{\mu \nu }^{a}F_{\mu \nu }^{a}+\frac{1}{2\alpha }\int
d^{4}x\left( \partial A^{a}\right) ^{2}+S_{\sigma }\right) }\mathcal{V}(%
\widetilde{\Omega })\;,\;  \label{m8}
\end{equation}
with the factor $\mathcal{V}(\widetilde{\Omega })$ given in
eqs.$\left( \ref {g3}\right) $,$\left( \ref{g4}\right) $. To
discuss the gluon propagator we proceed as before and retain only
the quadratic terms in expression $\left(
\ref{m8}\right) $ which contribute to the two-point correlation function $%
\left\langle A_{\mu }^{a}(k)A_{\nu }^{b}(-k)\right\rangle $. Expanding
around the nonvanishing vacuum expectation value of the auxiliary field, $%
\left\langle \sigma \right\rangle \neq 0$, one easily get
\begin{eqnarray}
\mathcal{Z}_{\mathrm{quadr}} &=&\mathcal{N}\int DA\frac{d\eta }{2\pi i\eta }%
e^{\eta (1-\rho (0,A))}e^{-\left( \frac{1}{4}\int d^{4}x(\left( \partial
_{\mu }A_{\nu }^{a}-\partial _{\mu }A_{\nu }^{a}\right) ^{2}+\frac{1}{%
2\alpha }\int d^{4}x\left( \partial A^{a}\right) ^{2}+\frac{1}{2}m^{2}\int
d^{4}x\left( A_{\mu }^{a}A_{\mu }^{a}\right) \right) }\;\;  \nonumber \\
&=&\mathcal{N}\int DA\frac{d\eta }{2\pi i}e^{\eta -\log \eta }e^{-\frac{1}{2}%
\sum_{q}A_{\mu }^{a}(q)\mathcal{Q}_{\mu \nu }^{ab}A_{\nu }^{b}(-q)}\;,
\label{m9}
\end{eqnarray}
where the factor $\mathcal{Q}_{\mu \nu }^{ab}$ is now given by
\begin{equation}
\mathcal{Q}_{\mu \nu }^{ab}=\left( \left( q^{2}+m^{2}+\frac{\eta Ng^{2}}{%
N^{2}-1}\frac{1}{2V}\frac{1}{q^{2}}\right) \delta _{\mu \nu }-q_{\mu }q_{\nu
}\left( \left( 1-\frac{1}{\alpha }\right) +\frac{\eta Ng^{2}}{N^{2}-1}\frac{1%
}{2V}\frac{1}{q^{4}}\right) \right) \delta ^{ab}\;.  \label{m10}
\end{equation}
Integrating over the gauge field, one has
\begin{equation}
\mathcal{Z}_{\mathrm{quadr}}=\mathcal{N}\int \frac{d\eta }{2\pi i}e^{\eta
-\log \eta }\left( \det \mathcal{Q}_{\mu \nu }^{ab}\right) ^{-\frac{1}{2}}=%
\mathcal{N}^{\prime }\int \frac{d\eta }{2\pi i}e^{f(\eta )}\;,  \label{m11}
\end{equation}
with
\begin{equation}
f(\eta )=\eta -\log \eta -\frac{3}{2}(N^{2}-1)\sum_{q}\log \left(
q^{4}+m^{2}q^{2}+\frac{\eta Ng^{2}}{N^{2}-1}\frac{1}{2V}\right) \;.
\label{m12}
\end{equation}
Evaluating $\mathcal{Z}_{\mathrm{quadr}}$ at the saddle point, yields
\begin{equation}
\mathcal{Z}_{\mathrm{quadr}}\approx e^{f(\eta _{0})}\;,  \label{m13}
\end{equation}
where $\eta _{0}$ is determined by the minimum condition
\begin{equation}
1-\frac{1}{\eta _{0}}-\frac{3}{4}\frac{Ng^{2}}{V}\sum_{q}\frac{1}{\left(
q^{4}+m^{2}q^{2}+\frac{\eta _{0}Ng^{2}}{N^{2}-1}\frac{1}{2V}\right) }=0\;.
\label{m14}
\end{equation}
Taking the thermodynamic limit, $V\rightarrow \infty $, and introducing the
Gribov parameter
\begin{equation}
\gamma ^{4}=\frac{\eta _{0}Ng^{2}}{N^{2}-1}\frac{1}{2V}\;\;,\;\;\;\;V%
\rightarrow \infty \;,  \label{m15}
\end{equation}
we get the gap equation in the presence of the dynamical gluon mass,
corresponding to a nonvanishing condensate $\left\langle A_{\mu }^{a}A_{\mu
}^{a}\right\rangle $, namely
\begin{equation}
\frac{3}{4}Ng^{2}\int \frac{d^{4}q}{\left( 2\pi \right) ^{4}}\frac{1}{%
q^{4}+m^{2}q^{2}+\gamma ^{4}}=1\;.  \label{m16}
\end{equation}
Note that the dynamical mass $m$ appears now explicitly in the gap
equation $\left( \mathrm{{\ref{m16}}}\right) $. To obtain the
gauge propagator, one goes back to the expression $\left(
\ref{m9}\right) $ which, when evaluated at the saddle point value
$\eta =\eta _{0}$, yields
\begin{equation}
\mathcal{Z}_{\mathrm{quadr}}=\mathcal{N}\int DAe^{-\frac{1}{2}\sum_{q}A_{\mu
}^{a}(q)\mathcal{Q}_{\mu \nu }^{ab}A_{\nu }^{b}(-q)}\;,  \label{m17}
\end{equation}
with
\begin{equation}
\mathcal{Q}_{\mu \nu }^{ab}=\left( \left( q^{2}+m^{2}+\frac{\gamma ^{4}}{%
q^{2}}\right) \left( \delta _{\mu \nu }-\frac{q_{\mu }q_{\nu }}{q^{2}}%
\right) +\frac{q_{\mu }q_{\nu }}{q^{2}}\left( \frac{q^{2}}{\alpha }%
+m^{2}\right) \right) \delta ^{ab}\;.  \label{m18}
\end{equation}
Thus, for the gauge propagator in the presence of the dynamical gluon mass $m
$ we get
\begin{equation}
\left\langle A_{\mu }^{a}(q)A_{\nu }^{b}(-q)\right\rangle =\delta
^{ab}\left( \frac{q^{2}}{q^{4}+m^{2}q^{2}+\gamma ^{4}}\left( \delta _{\mu
\nu }-\frac{q_{\mu }q_{\nu }}{q^{2}}\right) +\frac{\alpha }{q^{2}+\alpha
m^{2}}\frac{q_{\mu }q_{\nu }}{q^{2}}\right) \;.  \label{m19}
\end{equation}
We note that, due to the presence of the mass $m$, the infrared
suppression of the transverse component of the gluon propagator is
enforced.\ Moreover, also the longitudinal component gets
suppressed.

\subsection{The infrared behavior of $\mathcal{G}_{tr}(k)$ in the presence
of $\left\langle A_{\mu }^{a}A_{\mu }^{a}\right\rangle $}

It remains now to discuss the infrared behavior of the Green function $%
\mathcal{G}_{tr}(k)$ in the presence of $\left\langle A_{\mu }^{a}A_{\mu
}^{a}\right\rangle $. This can be easily worked out by repeating the
analysis done in the previous sections. From the expression of the gluon
propagator $\left( \mathrm{{\ref{m19}}}\right) $, it follows that
\begin{equation}
\mathcal{G}_{tr}(k)\approx \frac{1}{k^{2}}\frac{1}{1-\rho (k)}\;,
\label{gg1}
\end{equation}
with
\begin{eqnarray}
\rho (k) &=&g^{2}\frac{N}{N^{2}-1}\frac{k_{\mu }k_{\nu }}{k^{2}}\int \frac{%
d^{4}q}{\left( 2\pi \right) ^{4}}\frac{1}{\left( k-q\right) ^{2}}%
\left\langle A_{\mu }^{Ta}(q)A_{\nu }^{Ta}(-q)\right\rangle \;  \nonumber \\
&=&g^{2}N\frac{k_{\mu }k_{\nu }}{k^{2}}\int \frac{d^{4}q}{\left( 2\pi
\right) ^{4}}\frac{1}{\left( k-q\right) ^{2}}\frac{q^{2}}{%
q^{4}+m^{2}q^{2}+\gamma ^{4}}\left( \delta _{\mu \nu }-\frac{q_{\mu }q_{\nu }%
}{q^{2}}\right) \;.  \label{gg2}
\end{eqnarray}
Also, from the gap equation $\left( \mathrm{{\ref{m16}}}\right) $, one has
\begin{equation}
Ng^{2}\int \frac{d^{4}q}{\left( 2\pi \right) ^{4}}\frac{1}{%
q^{4}+m^{2}q^{2}+\gamma ^{4}}\left( \delta _{\mu \nu }-\frac{q_{\mu }q_{\nu }%
}{q^{2}}\right) =\delta _{\mu \nu }\;,  \label{gg3}
\end{equation}
so that
\begin{eqnarray}
1-\rho (k) &=&Ng^{2}\frac{k_{\mu }k_{\nu }}{k^{2}}\int \frac{d^{4}q}{\left(
2\pi \right) ^{4}}\frac{k^{2}-2qk}{\left( k-q\right) ^{2}}\frac{1}{%
q^{4}+m^{2}q^{2}+\gamma ^{4}}\left( \delta _{\mu \nu }-\frac{q_{\mu }q_{\nu }%
}{q^{2}}\right) \;.  \nonumber \\
&&  \label{gg4}
\end{eqnarray}
Thus, for $k\approx 0$,
\begin{equation}
\left( 1-\rho (k)\right) _{k\approx 0}\approx \frac{3Ng^{2}\mathcal{J}}{4}%
k^{2}\;,  \label{gg5}
\end{equation}
where $\mathcal{J}$ stands for the value of the integral
\begin{equation}
\mathcal{J=}\int \frac{d^{4}q}{\left( 2\pi \right) ^{4}}\frac{1}{%
q^{2}(q^{4}+m^{2}q^{2}+\gamma ^{4})}\;,  \label{gg6}
\end{equation}
which is ultraviolet finite. Therefore, for the Green function
$\mathcal{G}_{tr}(k)$, we get
\begin{equation}
\left( \mathcal{G}_{tr}(k)\right) _{k\approx 0}\approx \frac{4}{3Ng^{2}%
\mathcal{J}}\frac{1}{k^{4}}\;,  \label{gg7}
\end{equation}
exhibiting the infrared enhancement which, thanks to the gap equation $%
\left( \mathrm{{\ref{m16}}}\right) $, turns out to hold also in
the presence of the gluon condensate $\left\langle A_{\mu
}^{a}A_{\mu }^{a}\right\rangle $.

\section{Comparison with the results obtained from lattice simulations and
from the Schwinger-Dyson equations}

Having investigated the infrared behavior of the gluon propagator
and of the Green function $\left( \mathcal{G}_{tr}(k)\right) $, as
summarized by equations $\left( \mathrm{{\ref{m19}}}\right) $ and
$\left( \mathrm{{\ref {gg7}}}\right) $, it is useful to make a
comparison with the results already available from lattice
simulations and from the studies of the Schwinger-Dysons
equations. Let us begin with the lattice data

\subsection{Comparison with the lattice data}

In a series of papers
\cite{Giusti:1996kf,Giusti:1999im,Giusti:2000yc}, Giusti et al.
have managed to put the linear covariant gauges on the lattice.
This has allowed for a numerical investigation of the transverse
as well as of the longitudinal component of the gluon propagator.
Following \cite{Giusti:2000yc}, let us introduce the transverse
and longitudinal form factors $D_{T}(q)$ and $D_{L}(q)$ through
\begin{equation}
\left\langle A_{\mu }^{a}(q)A_{\nu }^{b}(-q)\right\rangle =\delta
^{ab}\left( \frac{D_{T}(q)}{q^{2}}\left( \delta _{\mu \nu }-\frac{q_{\mu
}q_{\nu }}{q^{2}}\right) +\frac{D_{L}(q)}{q^{2}}\frac{q_{\mu }q_{\nu }}{q^{2}%
}\right) \;.  \label{d1}
\end{equation}
The results obtained in
\cite{Giusti:1996kf,Giusti:1999im,Giusti:2000yc} show that both
$D_{T}(q)$ and $D_{L}(q) $ are suppressed in the low momentum
region, see for instance Fig.3 and Fig.4 of \cite{Giusti:2000yc}.
Our results are in qualitative agreement with
the lattice data. Indeed, from the expression $\left( \mathrm{{\ref{m19}}}%
\right) $, we obtain
\begin{eqnarray}
D_{T}(q) &=&\frac{q^{4}}{q^{4}+m^{2}q^{2}+\gamma ^{4}}\;,  \nonumber \\
D_{L}(q) &=&\frac{\alpha q^{2}}{q^{2}+\alpha m^{2}}\;,  \label{d2}
\end{eqnarray}
exhibiting infrared suppression. Note that, at least within the
approximation considered in the present work, the suppression of
the longitudinal form factor $D_{L}(q)$ in eq.$\left(
\mathrm{{\ref{d2}}}\right)
$ is a consequence of the dynamical gluon mass, due to the gluon condensate $%
\left\langle A_{\mu }^{a}A_{\mu }^{a}\right\rangle $, as already
pointed out in \cite{Dudal:2003by}. Concerning now the ghost
propagator and the Green function $\mathcal{G}_{tr}(k)$, to our
knowledge, no results from lattice data are available so far.

\subsection{Comparison with the results obtained from the Schwinger-Dysons
equations }

The infrared behavior of the gluon and ghost propagator has been
investigated within the Schwinger-Dyson framework in
\cite{Alkofer:2003jr}. Here, a power-law \textit{Ansatz } for the
transverse and longitudinal form factors of the gluon propagator
as well as for the ghost form factor $D_{gh}(q)$ has been
employed, according to
\begin{eqnarray}
D_{T}(q) &\approx &\left( q^{2}\right) ^{\sigma }\;,  \nonumber \\
D_{L}(q) &\approx &\left( q^{2}\right) ^{\rho }\;,  \label{sd1}
\end{eqnarray}
and
\begin{eqnarray}
\left\langle \overline{c}^{a}(q)c^{b}(-q)\right\rangle  &=&\delta ^{ab}\frac{%
D_{gh}(q)}{q^{2}}\;,  \nonumber \\
D_{gh}(q) &\approx &\left( q^{2}\right) ^{\beta }\;,  \label{sd2}
\end{eqnarray}
The results obtained for the infrared exponents $(\sigma, \rho,
\beta)$ turn out to be similar to those of the Landau gauge,
namely\footnote{The explicit values of these infrared exponents as
well as their dependence from the gauge parameter can be found in
\cite{Alkofer:2003jr}.}
\begin{eqnarray}
\sigma  &>&0\;,  \nonumber \\
\rho  &>&0\;,  \nonumber \\
-\beta  &=&\frac{\sigma }{2}=\frac{\rho }{2}\;,  \label{sd3}
\end{eqnarray}
indicating an infrared suppression of the transverse and
longitudinal gluon form factors, and an infrared enhancement of
the ghost propagator. Concerning the gluon propagator, these
results are in qualitative agreement with our results as well as
with the lattice data. However, concerning the ghost propagator,
we have found that, instead of the ghost form factor, the quantity
which is enhanced in the infrared is $\mathcal{G}_{tr}(k)$. For a
better understanding of this point, it is worth reminding here
that the result for the infrared exponents in eq.$\left(
\mathrm{{\ref{sd3}}}\right) $ has been obtained by using a
bare-vertex truncation scheme \cite{Alkofer:2003jr}. This
approximation has been proven successful in the Landau gauge
\cite{vonSmekal:1997is,Atkinson:1997tu,Atkinson:1998zc,Watson:2001yv,Zwanziger:2001kw,Lerche:2002ep}.
In particular, in the Landau gauge, no qualitative difference has
been found if bare vertices are replaced by vertices dressed
according to the Slanov-Taylor identities. This feature of the
Landau gauge is believed to be deeply related to the
nonrenormalization theorem of the ghost-antighost-gluon vertex,
which holds to all orders of perturbation theory
\cite{Blasi:1990xz,Piguet:1995er}. Recently, the
nonrenormalization theorem of the ghost-antighost-gluon vertex in
the Landau gauge has been investigated through lattice simulations
in \cite{Cucchieri:2004sq}, which have provided indications of its
validity beyond perturbation theory. However, to our knowledge, no
such a theorem is available in linear covariant gauges, for a
nonvanishing value of the gauge parameter $\alpha$. Furthermore,
according to the authors \cite{Alkofer:2003jr}, it is yet an open
question whether the values of the infrared exponents in
eq.$\left(\mathrm{{\ref{sd3}}}\right)$ remain unchanged if bare
vertices are replaced by dressed ones. Our results suggest that a
different behavior might be expected when dressed vertices would
be employed.

\section{Conclusion}

In this work we have attempted at analyzing the effects of the
Gribov copies on the gluon propagator in linear covariant gauges.
By considering small values of the gauge parameter $\alpha$, a few
properties of the Gribov copies have been established, allowing us
to investigate the infrared behavior of the gluon two-point
function.
\\\\As in the case of the Landau gauge, it turns out that the transverse component
of the gluon propagator is suppressed in the infrared. Moreover,
the longitudinal part is left unchanged, as shown in eq.$\left(
\mathrm{{\ref{p15}}}\right)$. The infrared behavior of the gluon
propagator has been investigated also in the presence of the gluon
condensate $\left\langle A_{\mu }^{a}A_{\mu }^{a}\right\rangle $.
In this case, the infrared suppression of the transverse component
is enforced. Furthermore, its longitudinal component turns out to
be suppressed as well, as expressed by eq.$\left(
\mathrm{{\ref{m19}}}\right) $. These results are in qualitative
agreement with those obtained from lattice simulations and from
the analysis of the Schwinger-Dyson equations. \\\\
Concerning now the behavior of the ghost fields in linear
covariant gauges, the output of our analysis is that, instead of
the ghost propagator, the Green function which exhibits infrared
enhancement is given by $\mathcal{G}_{tr}(k)$, as defined in
eq.$\left( \mathrm{{\ref{gtr}}}\right)$. It should be remarked
that $\mathcal{G}_{tr}(k)$ does not coincide with the ghost
propagator for a generic value of the gauge parameter $\alpha $.
However, $\mathcal{G}_{tr}(k) $ reduces to the ghost two-point
function for vanishing $\alpha $, so that our results turn out to
coincide with those of the Landau gauge in the limit $\alpha
\rightarrow 0$.
\\\\Needless to say, many aspects of the covariant linear gauges
remain still to be investigated. A partial list of
them is: \\\\
As already pointed out, a suitable auxiliary functional
corresponding to the linear covariant gauge fixing condition is
not yet at our disposal. As in the case of the Landau gauge
\cite{Dell'Antonio:1989jn,Dell'Antonio:1991xt,Semenov}, this
functional could be very helpful for a characterization of the
properties of the Gribov
copies not attainable by infinitesimal gauge transformations. \\\\
From the present analysis, it emerges that the Green function
$\mathcal{G}_{tr}(k)$ has a special role, as it obeys the
Gribov-Zwanziger horizon condition and reduces to the ghost
two-point function in the Landau gauge. Although its dependence
from the transverse component $A^{aT}_{\mu}$ of the gauge field
suggests that it might have a deeper meaning, it would be worth to
have a better understanding of $\mathcal{G}_{tr}(k)$.  \\\\
It would be useful to have a consistent framework to compute
quantum corrections to the gluon propagator and to
$\mathcal{G}_{tr}(k)$. This would amount to construct a local
renormalizable action in linear covariant gauges incorporating the
effects of the Gribov copies, as done by Zwanziger in the Landau
gauge \cite{Zwanziger:1989mf,Zwanziger:1992qr}. \\\\
Finally, it would be interesting to have more data in the linear
covariant gauges from lattice simulations on the gluon and ghost
propagators as well as on the Green function
$\mathcal{G}_{tr}(k)$.

\section*{Acknowledgments.}
The Conselho Nacional de Desenvolvimento Cient\'{i}fico e
Tecnol\'{o}gico (CNPq-Brazil), the Faperj, Funda{\c c}{\~a}o de
Amparo {\`a} Pesquisa do Estado do Rio de Janeiro,
the SR2-UERJ and the Coordena{\c{c}}{\~{a}}o de Aperfei{\c{c}}%
oamento de Pessoal de N{\'{i}}vel Superior (CAPES) are gratefully
acknowledged for financial support.

\end{document}